\documentstyle[12pt,aaspp4]{article}
\begin{document}

\title{ The Discovery of a Luminous $z=5.80$ Quasar from the Sloan Digital Sky Survey$^1$}

\author{Xiaohui Fan\altaffilmark{\ref{Princeton}},
Richard L. White\altaffilmark{\ref{STScI}},
Marc Davis\altaffilmark{\ref{UCB}},
Robert H. Becker\altaffilmark{\ref{UCDavis},\ref{IGPP}},
Michael A. Strauss\altaffilmark{\ref{Princeton}},
Zoltan Haiman\altaffilmark{\ref{Princeton}},
Donald P. Schneider\altaffilmark{\ref{PennState}},
Michael D. Gregg\altaffilmark{\ref{UCDavis},\ref{IGPP}},
James E. Gunn\altaffilmark{\ref{Princeton}},
Gillian R. Knapp\altaffilmark{\ref{Princeton}},
Robert H. Lupton\altaffilmark{\ref{Princeton}},
John E. Anderson, Jr.\altaffilmark{\ref{Fermilab}},
Scott F. Anderson\altaffilmark{\ref{Washington}}, 
James Annis\altaffilmark{\ref{Fermilab}},
Neta A. Bahcall\altaffilmark{\ref{Princeton}},
William N. Boroski\altaffilmark{\ref{Fermilab}},
Robert J. Brunner\altaffilmark{\ref{Caltech}},
Bing Chen\altaffilmark{\ref{JHU}},
Andrew J. Connolly\altaffilmark{\ref{Pitt}}, 
Istvan Csabai\altaffilmark{\ref{JHU}},
Mamoru Doi\altaffilmark{\ref{UTokyo}},
Masataka Fukugita\altaffilmark{\ref{CosmicRay},\ref{IAS}},
G. S. Hennessy\altaffilmark{\ref{USNO}},
Robert B. Hindsley\altaffilmark{\ref{NRL}},
Takashi Ichikawa\altaffilmark{\ref{Tohoku}},
\v{Z}eljko Ivezi\'{c}\altaffilmark{\ref{Princeton}},
Avery Meiksin\altaffilmark{\ref{Edinburgh}},
Timothy A. McKay\altaffilmark{\ref{Michigan}},
Jeffrey A. Munn\altaffilmark{\ref{Flagstaff}},
Heidi J. Newberg\altaffilmark{\ref{RPI}},
Robert Nichol\altaffilmark{\ref{CMU}},
Sadanori Okamura\altaffilmark{\ref{UTokyo}},
Jeffrey R. Pier\altaffilmark{\ref{Flagstaff}},
Maki Sekiguchi\altaffilmark{\ref{CosmicRay}},
Kazuhiro Shimasaku\altaffilmark{\ref{UTokyo}},
Alexander S. Szalay\altaffilmark{\ref{JHU}},
Gyula P. Szokoly\altaffilmark{\ref{Potsdam}},
Aniruddha R. Thakar\altaffilmark{\ref{JHU}},
Michael S. Vogeley\altaffilmark{\ref{Drexel}},
Donald G. York\altaffilmark{\ref{Chicago}}}

\altaffiltext{1}{Based on observations obtained with the
Sloan Digital Sky Survey, 
which is owned and operated by the Astrophysical
Research Consortium,
and at the W. M. Keck Observatory, which is operated as a scientific
partnership among the California Institute of Technology,
the University of California, and NASA,
and was made possible by the generous financial support of the W. M. Keck
Foundation.}
\newcounter{address}
\setcounter{address}{2}
\altaffiltext{\theaddress}{Princeton University Observatory, Princeton,
NJ 08544
\label{Princeton}}
\addtocounter{address}{1}
\altaffiltext{\theaddress}{Space Telescope Science Institute, Baltimore, MD 21
218
\label{STScI}}
\addtocounter{address}{1}
\altaffiltext{\theaddress}{Department of Astronomy, University of California,
Berkeley, CA 94720-3411
\label{UCB}}
\addtocounter{address}{1}
\altaffiltext{\theaddress}{Physics Department, University of California, Davis, CA
95616
\label{UCDavis}}
\addtocounter{address}{1}
\altaffiltext{\theaddress}{IGPP/Lawrence Livermore National Laboratory, Livermore,
CA 95616
\label{IGPP}}
\addtocounter{address}{1}
\altaffiltext{\theaddress}{Department of Astronomy and Astrophysics,
The Pennsylvania State University,
University Park, PA 16802
\label{PennState}}
\addtocounter{address}{1}
\altaffiltext{\theaddress}{Fermi National Accelerator Laboratory, P.O. Box 500,
Batavia, IL 60510
\label{Fermilab}}
\addtocounter{address}{1}
\altaffiltext{\theaddress}{University of Washington, Department of Astronomy,
Box 351580, Seattle, WA 98195
\label{Washington}}
\addtocounter{address}{1}
\altaffiltext{\theaddress}{
Department of Astronomy, California Institute of Technology,
Pasadena, CA 91125
\label{Caltech}}
\addtocounter{address}{1}
\altaffiltext{\theaddress}{
Department of Physics and Astronomy, The Johns Hopkins University,
   3701 San Martin Drive, Baltimore, MD 21218, USA
\label{JHU}}
\addtocounter{address}{1}
\altaffiltext{\theaddress}{
Department of Physics and Astronomy, University of Pittsburgh, Pittsburgh, PA 15260
\label{Pitt}}
\addtocounter{address}{1}
\altaffiltext{\theaddress}{Department of Astronomy and Research Center 
  for the Early Universe, School of Science, University of Tokyo, Hongo,
  Bunkyo, Tokyo, 113-0033, Japan
\label{UTokyo}}
\addtocounter{address}{1}
\altaffiltext{\theaddress}{Institute for Cosmic Ray Research, University of
Tokyo, Midori, Tanashi, Tokyo 188-8502, Japan
\label{CosmicRay}}
\addtocounter{address}{1}
\altaffiltext{\theaddress}{Institute for Advanced Study, Olden Lane,
Princeton, NJ 08540
\label{IAS}}
\addtocounter{address}{1}
\altaffiltext{\theaddress}{U.S. Naval Observatory,
3450 Massachusetts Ave., NW,
Washington, DC  20392-5420
\label{USNO}}
\addtocounter{address}{1}
\altaffiltext{\theaddress}{Remote Sensing Division, Code 7215, Naval
  Research Laboratory, 4555 Overlook Ave. SW, Washington, DC 20375
\label{NRL}}
\addtocounter{address}{1}
\altaffiltext{\theaddress}{Astronomical Institute,
Tohoku University,
Aoba, Sendai 980-8578
Japan
\label{Tohoku}}
\addtocounter{address}{1}
\altaffiltext{\theaddress}{
Institute for Astronomy, University of Edinburgh, Edinburgh EH9 3HJ, UK
\label{Edinburgh}}
\addtocounter{address}{1}
\altaffiltext{\theaddress}{University of Michigan, Department of Physics,
        500 East University, Ann Arbor, MI 48109
\label{Michigan}}
\addtocounter{address}{1}
\altaffiltext{\theaddress}{U.S. Naval Observatory, Flagstaff Station,
P.O. Box 1149,
Flagstaff, AZ  86002-1149
\label{Flagstaff}}
\addtocounter{address}{1}
\altaffiltext{\theaddress}{Rensselaer Polytechnic Insitute, Dept. of Physics, Applied Physics, and Astronomy, Troy, NY 12180
\label{RPI}}
\addtocounter{address}{1}
\altaffiltext{\theaddress}{
Department of Physics, Carnegie Mellon University, Pittsburgh, PA 15213
\label{CMU}}
\addtocounter{address}{1}
\altaffiltext{\theaddress}{
Astrophysikalisches Institut Potsdam
An der Sternwarte 16 D-14482 Potsdam, Germany
\label{Potsdam}}
\addtocounter{address}{1}
\altaffiltext{\theaddress}{Department of Physics, Drexel University,
  3141 Chestnut St., Philadelphia, PA 19104
\label{Drexel}}
\addtocounter{address}{1}
\altaffiltext{\theaddress}{University of Chicago, Astronomy \& Astrophysics
Center, 5640 S. Ellis Ave., Chicago, IL 60637
\label{Chicago}}

\begin{abstract}
We present observations of SDSSp J104433.04--012502.2, a luminous quasar at
$z=5.80$ discovered from Sloan Digital Sky Survey (SDSS) multicolor
imaging data. This object was selected as an $i'$-band dropout object,
with $i^*=21.8 \pm 0.2$, $z^*=19.2 \pm 0.1$. It has an absolute magnitude 
$M_{1450} = -27.2$ ($H_{0} =50$ km s$^{-1}$ Mpc$^{-1}$, $q_{0} = 0.5$). 
The spectrum shows a strong and broad Ly$\alpha$ emission line,
strong Ly$\alpha$ forest absorption lines with a mean continuum decrement
$D_{A} = 0.91$,  and a Lyman Limit System at $z=5.72$. 
The spectrum also shows strong OI and SiIV emission lines
similar to those of quasars at $z\lesssim 5$,
suggesting that these metals were produced at redshift beyond six.  
The lack of a Gunn-Peterson trough in the spectrum indicates
that the universe is already highly ionized at $z \sim 5.8$.
Using a high-resolution spectrum in the Ly$\alpha$ forest region, 
we place a conservative
upper limit of the optical depth due to the Gunn-Peterson effect of $\tau < 0.5$ in regions
of minimum absorption. 
The Ly$\alpha$ forest absorption in this object is much stronger than
that in quasars at $z\lesssim 5$.  The object is unresolved in a deep
image with excellent seeing, implying that it is unlensed. 
The black hole mass of this quasar is $\sim 3 \times 10^9 M_{\odot}$
if we assume that it is radiating at the Eddington luminosity and no lensing amplification,
implying that it resides in a very massive dark matter halo.
The discovery of one quasar at $M_{1450} < -27$ in a survey area of  600 deg$^2$ 
is consistent with an extrapolation of the observed luminosity function
at lower redshift.
The abundance and evolution of such quasars can provide sensitive tests  of
models of quasar and galaxy formation.
\end{abstract}

\section{Introduction}

At what epoch did the first generation of galaxies and quasars
form? How was the universe re-ionized, ending
the ``dark ages'' (\cite{Rees98})?
These fundamental questions can only be answered with studies of high-redshift objects.
The last few years have witnessed the first direct observations
of galaxies at redshift higher than five
(\cite{Dey98}, \cite{Weymann98}, \cite{Spinrad98}, \cite{Chen99},
\cite{Breugel99}, \cite{Hu99},
see also the review by \cite{Stern99}), while detailed studies of
the ensemble properties and large scale distribution of galaxies at $z\sim4$
have begun (Steidel et al.~1998, 1999).
Several quasars have been found at $z\gtrsim 5$ (Fan et al.~1999, 2000a,
\cite{Zheng00}), including a low-luminosity quasar at $z=5.50$ (\cite{Stern00}).

Studies of high-redshift quasars provide important probes of this critical
epoch in cosmic evolution. 
The lack of the Gunn-Peterson (1965) effect in the absorption spectrum of
a $z\sim 5.0$ quasar (\cite{S99}) indicates that the universe is
already highly ionized at that redshift. 
The exact epoch of re-ionization could be determined from the absorption
spectra of quasars at even higher redshift (\cite{Jordi97},
\cite{HL99}).
The study of the luminosity function of high-redshift quasars will constrain
models of quasar and galaxy evolution (\cite{HL98},
\cite{Haehnelt98}),  and
determine whether it was  UV radiation from AGNs or from young massive
stars that
re-ionized the universe, ending the ``dark ages'' (\cite{HL98}).
Measurements of the chemical abundance in the quasar environment 
will reveal the metal production process at the very early stage of
galaxy evolution (\cite{HF99}).
Finally, luminous high-redshift quasars represent high peaks of density fields,
and may be the markers of large scale structure at these early epochs 
(\cite{George99}, \cite{HH00}, \cite{MW00}).

The Sloan Digital Sky Survey (SDSS;
\cite{York00}) is using
a dedicated 2.5m telescope and a large format CCD camera (\cite{Gunnetal})
at the Apache Point Observatory in New Mexico
to obtain images in five broad bands ($u'$, $g'$, $r'$, $i'$ and $z'$,
centered at 3540, 4770, 6230, 7630 and 9130 \AA, respectively; \cite{F96}) 
over 10,000 deg$^2$ of high Galactic latitude sky.
The multicolor data from SDSS have proven to be very effective in selecting
high-redshift quasars: more than 50 quasars at $z>3.5$ have been
discovered to date from about 600 deg$^2$ of imaging data
(Fan et al.~1999, 2000a, \cite{HET1}, \cite{Zheng00}).
The inclusion of the reddest band, $z'$, in principle enables the
detection of quasars up to $ z \sim 6.5$ in SDSS data.

In this paper, we report the discovery of SDSSp J104433.04--012502.2
(the name reflecting its
J2000 coordinates from the preliminary SDSS astrometry, 
accurate to $\sim 0.1''$ in each coordinate), a very luminous, 
``$i'$-dropout''  quasar at $z=5.80$, selected by its very red $i^*-z^*$ color.
In a $\Lambda$-dominated flat universe ($H_{0}$ =  65 km s$^{-1}$ Mpc$^{-1}$,
$\Lambda=0.65$ and $\Omega=0.35$, referred to as the $\Lambda$-model in 
this paper, \cite{OS95}, \cite{KT95}), $z=5.80$ corresponds to an age of
0.9 Gyr in an universe 13.9 Gyr old, or a look-back time of 93.2\%
of the age of the universe.
Similarly, the universe was 0.7 Gyr old at $z=5.80$ in a universe
13.0 Gyr old at present for a model with 
$\Omega=1$ and $H_{0}$ = 50 km s$^{-1}$ Mpc$^{-1}$, which we refer to as
the $\Omega=1$ model in this paper. 
We present the photometric observations and target selection in \S 2,
and the spectroscopic observations in \S 3.  
In \S 4, we discuss the cosmological implications,
including the constraints on the
Gunn-Peterson effect, quasar evolution models, and black hole formation.

\section{Photometric Observation and Target Selection}

The object SDSSp J104433.04--012502.2 (hereafter SDSS 1044--0125 for
brevity) was selected from the SDSS imaging data based on
its extremely red $i^* - z^*$ color. 
The photometry of this object is summarized in Table 1.
The photometric observations of this region were obtained by the 
SDSS imaging camera on 2000 March 4 during the
SDSS commissioning phase.
For off-equator scans, 
the telescope 
moves along a great circle, with
the photometric camera drift-scanning at the sidereal rate.
The effective exposure time is 54.1 seconds in each band.
The seeing in the $i'$ and $z'$ bands was about $1.9''$.
The photometric calibration is provided by an auxiliary 20-inch
telescope at the same site (Uomoto et al.~2000, in preparation).
The photometric zeropoint is accurate to about 7\% in $u'$ and $z'$ and
2\% in $g'$, $r'$ and $i'$. 
Because the definition of the photometric system is not yet finalized,
we quote measured magnitudes using asterisks (e.g., $i^*$) 
to represent preliminary
photometry, while referring  to the filters with primes (e.g., $i'$).
SDSS 1044--0125 is undetected in $u'$, $g'$ and $r'$. 
The object is detected at the 5-$\sigma$ level in the $i'$ band.
The $z'$ detection is of very high significance, with
$i^* - z^* = 2.58 \pm 0.20$. 
Data are quoted as asinh magnitudes (\cite{Luptitude}) and
are on the AB magnitude system (\cite{F96}).
Finding charts for SDSS 1044--0125 in the $i'$ and $z'$ bands are shown
in Figure~1.

As shown by \cite{F99}, the colors of high-redshift quasars are strong
functions of redshift in the SDSS filter system,
as first the Ly$\alpha$ forest and then the Lyman Limit Systems 
move through the filter system.
At $z>3.6$, quasars become very red in $g^*-r^*$ while remaining blue 
in $r^*-i^*$, and can be
readily distinguished from stars based on these colors.
At $z>4.6$, quasars become very red in $r^*-i^*$. 
They are easily distinguished from red stars in the 
$r^*-i^*$ vs. $i^*-z^*$ color-color diagram (Fan et al.~1999, 2000a).
Quasars at $z\sim 5.5$ have very similar $r^*-i^*$ and  $i^*-z^*$
colors to those of late M stars.
Finally, at $z\gtrsim 5.7$, the Ly$\alpha$ emission line begins to move out of 
the SDSS $i'$ filter.
With a predicted $i^*-z^* \gtrsim 2$, quasars at  such redshifts become 
$i'$-band dropout objects.
However, unlike the lower-redshift $g'$ and $r'$-dropout quasars,
these quasars have only one measurable optical color,
as they  will be completely undetected in $r'$.
The SDSS alone cannot provide a constraint on the quasar's continuum shape
redward of the Ly$\alpha$ emission. It is thus difficult to
distinguish them from other classes of red  objects
without additional information, such as near-infrared photometry (see also
Zheng et al.~2000) or detection in the radio or X-ray.

In fact, the only other known class of stellar objects with such red colors
($i^*-z^* \gtrsim 2$) are the extremely cool stars/substellar objects
with spectral type L or T
(\cite{Kirk99}, \cite{Strauss99}, \cite{T00}, \cite{Fan00b}, \cite{Leggett00}).
Most of these objects are brown dwarfs with mass below the hydrogen
burning limit. 
Although L and T dwarfs are very rare on the sky 
(the density for L dwarfs is 1 per $\sim 15$ deg$^2$ for $i^* < 20$, 
\cite{Fan00b}),
they are  still more numerous than are $z > 5$ quasars, 
and are the major contaminants of searches for $i'$-dropout quasars.
Near infrared photometry can be used to  separate  high-redshift quasar 
candidates from these cool dwarfs:
the continuum shape of quasars is relatively flat towards near-IR bands,
while the flux of L dwarfs continues to rise sharply
towards longer wavelengths. 

We have selected $i'$-dropout candidates with $i^*-z^* \gtrsim 2$ and
$z^* < 19.5$ from about 600 deg$^2$ of SDSS imaging data.
These 600 deg$^2$ overlap the publicly released area of
the Two Micron All Sky Survey (2MASS). 
SDSS 1044--0125 is the only $i'$-dropout source in the 2MASS covered area
that is {\em not detected} at the 7-$\sigma$ level in any band in the
2MASS Point Source Catalog. All other sources are detected in at least
one band at more than 10 $\sigma$ (e.g. \cite{Fan00b}, \cite{Leggett00}).

Is the non-detection in 2MASS sufficient to rule out the possibility
that SDSS 1044--0125 is a brown dwarf?
In this area of the sky, the 2MASS Point Source Catalog has  7-$\sigma$ limiting magnitudes
of roughly $J=16.7$, $H=15.9$ and $K_s=15.0$. 
This indicates a 7-$\sigma$ upper limit on the optical-IR colors of
SDSS 1044--0125: $z^* - J < 2.5$, $z^*-H < 3.3$ and $z^*-K_s < 4.2$.
There are three L dwarfs in Fan et al.~(2000b) that have $i^*-z^* > 2$:
SDSS 0330--0025 ($i^*-z^* = 2.13$, spectral type L2), SDSS 0539--0059
($i^*-z^* = 2.31$, L5) and SDSS 1326--0038 ($i^*-z^*=2.61$, L8).
They all have $z^* - J \gtrsim 2.7$, $z^*-H \gtrsim 3.6$ and 
$z^* - K_s \gtrsim 4.2$ (see Tables 2 and 5 in \cite{Fan00b}).
Although the current L dwarf sample is small, 
the $i^* - z^*$ and the $z^* - J$ (or 
$H$, $K_s$) colors seem to be reasonable indicators of the spectral type 
for L dwarfs (\cite{Kirk99}, \cite{Fan00b}) 
Therefore, if SDSS 1044--0125 ($i^* - z^* = 2.6$) had been an L dwarf,
we would expect it to have been detected by 2MASS with high
significance. 
Thus the non-detection in the 2MASS passbands indicates that SDSS 1044--0125
is unlikely to be a L dwarf, but is rather a source with a much
flatter IR continuum, such as a quasar at $z>5.6$ or a compact galaxy
at $z>1$.

A $K'$-band image of SDSS 1044-0125 was obtained on the night of
2000 April 17, using the Near Infrared Camera (NIRC, Matthews \&
Soifer 1994) on the Keck~I telescope under photometric skies and good
seeing.  
The observations consisted of a nine-point dither pattern, integrating
for 3 $\times$ 10s coadds at each location.  The data were flattened,
sky-subtracted, shifted, and stacked using the DIMSUM package in IRAF.
The images show SDSS 1044-0125 to be an unresolved point source with
FWHM = $0\farcs375$.  Photometry through a 2\arcsec\ radius aperture
yields $K' = 17.02 \pm 0.04$, referenced to the standards of
Persson et al.~(1998).  There are no companions or associated
structure within 20\arcsec\ of the quasar to a $3\sigma$ point source limiting
magnitude of $K' \approx 22.3$.  

\section{Spectroscopy}

Low and high dispersion spectra of SDSS 1044--0125 were obtained
with the Echelle Spectrograph and Imager (ESI; \cite{ESI})
on the Keck II telescope on the night of 2000 April 6. 
The night was photometric with $0.9''$ seeing.
A 1200 second exposure was taken through a $0.7''$ slit 
in the low-dispersion mode of ESI. 
In this mode, a prism is used for dispersion.
The spectrum covers 3900 \AA\ to about 10000 \AA. 
The dispersion varies roughly linearly with wavelength from 0.8 \AA/pixel 
at 3900 \AA\ to 10 \AA/pixel at 10000 \AA.
In addition, two 1200 second high resolution spectra were taken
through a $1.0''$ slit in the
echellette mode of ESI. 
In this mode, the spectral range of 3900 \AA\ to 11000 \AA\ is
covered in ten spectral orders with a constant dispersion
of 11.4 km s$^{-1}$ pixel$^{-1}$. 
Wavelength calibrations were performed with observations of 
Hg-Ne-Xe lamps in the low-dispersion prism mode,
and a Cu-Ar lamp in the echellette mode.
The spectrophotometric
standard G191-B2B (\cite{Massey1}, \cite{Massey2}) was observed for
flux calibration. 
All observations were carried out at the  parallactic angle.
The data were reduced with standard IRAF routines. 

Figure 2 shows the final spectrum combining the low and high dispersion
observations (total exposure time of 3600s), over a wavelength range of
4500 - 10000 \AA, and binned to 4 \AA/pixel.
The telluric absorption bands were removed using the standard star spectrum.
The absolute flux scale of the spectrum is adjusted so that it 
reproduces the SDSS $z^*$ magnitude.
The signal-to-noise ratio at $\lambda > 8000$ \AA\ is 15 -- 20 per pixel.
The spectrum of SDSS 1044--0125 shows the unambiguous signature of 
a very  high redshift quasar:
the broad and strong Ly$\alpha$+NV emission line at $\lambda \sim 8300$ \AA\
with a sharp discontinuity to the blue side, due to the onset of very
strong Ly$\alpha$ forest absorption.
The flux level drops by a factor of $\sim 10$ from the red side to the blue
side of the Ly$\alpha$+NV emission.
A Ly$\beta$+OVI emission line is detected at
$\sim$ 7000 \AA, with an additional flux decrement due to the onset
of Ly$\beta$ forest absorption lines. 
The spectrum shows no detectable flux at $\lambda < 6100$ \AA\
because of the presence of a Lyman Limit System (see \S 4).
Redward of Ly$\alpha$, two additional emission lines,
OI+SiII$\lambda 1302$ and SiIV+OIV]$\lambda 1400$ are also clearly visible.
The synthetic $i^*-z^*$ color calculated from the spectrum in Figure 2 
is 2.5, consistent with the SDSS measurement.

The redshift determination of such an object is not
straightforward. The Ly$\alpha$ emission line 
is severely affected by the strong 
Ly$\alpha$ forest lines (as well as by possible internal absorption lines).
For $z \sim 4$ quasars, Schneider, Schmidt \& Gunn (1991) show that the peak of the Ly$\alpha$ emission line
is typically at rest-frame 1219 \AA, but with a large scatter. 
Another strong line, CIV$\lambda$ 1549, is out of the range of our spectrum.
OI+SiII$\lambda 1302$ and SiIV+OIV]$\lambda 1400$ are relatively weak.
We use the central wavelengths of these two blends from \cite{F91}.
Gaussian fits of these two lines are listed in Table 2.
Based on these fits, we adopt a redshift of 5.80 $\pm$ 0.02
for SDSS 1044--0125.
Objective measurement of the equivalent width of the Ly$\alpha$ line is difficult 
(see the discussion by Schneider, Schmidt \& Gunn 1991).
Using the continuum level determined from the red side of Ly$\alpha$ emission,
we find the rest-frame equivalent width of Ly$\alpha$ +NV is $\sim 26$ \AA.
However, this number is clearly an underestimate due to the extreme
Ly$\alpha$ absorption, maybe by a factor of two.
Figure 2 indicates that the expected line center of Ly$\alpha$ at $z=5.80$
is severely absorbed, and that the blue wing of the line has almost completely
disappeared.
The measurement of the Ly$\beta$+OVI line is difficult as well.
We smooth the spectrum of Figure 2 to estimate the continuum level, 
and find  an equivalent width of $\sim 30$\AA.
This value is also highly uncertain due to the 
difficulty in defining a ``continuum'' in the Ly$\alpha$ forest region. 

It is interesting to note that the strengths of the 
OI+SiII$\lambda 1302$ and SiIV+OIV]$\lambda 1400$ line blends
are comparable to those at much lower redshift:
the average rest frame equivalent widths of these two lines
are $3.2 \pm 0.4$ \AA\ and $8.1 \pm 0.6$ \AA\ in the sample of 
30 quasars at $z\sim 4$ in Schneider, Schmidt \& Gunn (1991),
compared to 3.4 and 7.0 \AA\ in the case of SDSS 1044--0125.
Previous studies have shown that quasar environments at $z\sim 4$
have roughly solar or higher metallicities (\cite{HF99}).
Although we cannot derive the metallicity of SDSS 1044--0125 
based on these emission line measurements,
the existence of strong lines suggests
that in this system, the metallicity is already quite high
at this redshift. Assuming the metals were produced in stellar evolution,
the initial starburst and chemical enrichment of the quasar environment
must have happened at a very early epoch. 

From the spectrum in Figure 2, we derive its continuum AB magnitude
at rest frame 1280 \AA,  $AB_{1280} = 19.28$, after correcting
for interstellar extinction ($E(B-V) = 0.054$, from the map
of \cite{Schlegel98}).
In the $\Lambda$-model (see \S 1), 
SDSS 1044--0125 has an absolute magnitude $M_{1280} = -27.41$.
Assuming a power law continuum with $f_{\nu} \propto \nu^{-0.5}$
we find $M_{1450} = -27.50$ and  $M_{B} = - 27.96$.
In the $\Omega=1$ model (\S 1) 
it has $M_{1280} = -27.15$,  $M_{1450} = -27.24$, and $M_{B} = - 27.70$.
In this cosmology, the nearby luminous quasar 3C273 has
$M_{B} = -27.0$.
SDSS 1044--0125 is a very luminous quasar, about twice as luminous as 3C 273
(assuming that it is not amplified by lensing or beaming). 

SDSS 1044--0125 is detected neither in the FIRST radio survey
(\cite{FIRST}) at the 1mJy level at 20cm wavelength, 
nor in the ROSAT All Sky Survey (\cite{ROSAT}), implying a 3-$\sigma$ upper
limit of $3 \times 10^{-13}$ ergs cm$^{-2}$ s$^{-1}$ in the 0.1 --
2.4 keV band.
These result is not unexpected; only a few $z>4$ quasars have observed
X-ray or radio fluxes above this value (\cite{KBS00}, \cite{Schmidt95}).
A deep exposure with Chandra or XMM is needed to determine its X-ray properties.

\section{Discussion}

\subsection{Absorption Properties and Gunn-Peterson Effect}

The high luminosity of SDSS 1044--0125 makes it an ideal object
for high signal-to-noise ratio
observations to study the intergalactic medium at high redshift.
In order to detect continuum break
caused by the  Lyman Limit System,
an edge filter with width of 40 \AA\ was convolved with the low resolution
spectrum in Figure 2. 
A strong peak at 6131 \AA\ is detected in the convolved spectrum,
indicating the existence of a Lyman Limit System at $z_{LLS} = 5.72$.
No flux is detected blueward of this break.
A Lyman Limit System is usually detected within 0.1 of the emission 
redshift in essentially all quasars at $z>4$ (\cite{SSG91}, \cite{APM}, \cite{Fan99}).

The most striking feature of the spectrum of SDSS 1044--0125
is the very strong absorption caused by Ly$\alpha$ forest lines.
However, the flux level in the Ly$\alpha$ forest region
never reaches zero.
It lacks the Gunn-Peterson (1965) trough that would exist in the spectrum of
a quasar at redshift higher than the re-ionization redshift (\cite{HL99}),
indicating that the intergalactic medium is already highly ionized at $z \sim 5.8$.
We estimate the average continuum decrements as: 
$D_{A,B} \equiv \left\langle 1 - f_\nu^{obs}/f_\nu^{con} \right\rangle $, where
$f_\nu^{obs}$ and $f_\nu^{con}$ are the observed and the unabsorbed
continuum fluxes of the quasar, and 
$D_{A}$ and $D_{B}$ measure the decrements in the region between
rest-frame Ly$\alpha$ and Ly$\beta$ ($\lambda = 1050 - 1170$ \AA) and
between Ly$\beta$ and the Lyman Limit ($\lambda = 920 - 1050$ \AA), 
respectively (\cite{OK82}).
The measurements of $D_{A}$ and $D_{B}$ require knowledge of
the continuum shape redward of Ly$\alpha$.
However, with $D_{A}$ approaching unity, the effect of different slopes
is quite small.
Assuming a power law continuum $\nu^{\alpha}$ with $\alpha = -0.5$, as
indicated in Figure 3, we obtain $D_{A} = 0.91$ and $D_{B}$ = 0.94. 
Using a slope of --1.0 only changes the $D_{A}$ and $D_{B}$ values
to 0.92 and 0.96, respectively.
We therefore adopt $D_{A} = 0.91 \pm 0.02$ and
$D_{B} = 0.95 \pm 0.02$.
These values are in close accordance with those from quasar RD J030117+002025
($z=5.50$, \cite{Stern00}) and from distant galaxies in the Hubble Deep Field
(\cite{Weymann98}), and are much higher than that of the $z=5.00$ quasar
SDSSp J033829.31+002156.3 ($D_{A} = 0.75$, \cite{S99}), suggesting
that the strong evolution of the strength of the Ly$\alpha$ forest at $z>5$,
$N(z) \propto (1+z)^{2.3-2.75}$, measured at redshifts below five continues
to a redshift of nearly six.   Assuming this number density evolution,
\cite{Zuo93} and \cite{F99} show that at $z\sim 5.8$, 
the expected average $D_{A}$ ranges from 0.8 to 0.9.


We further derive an upper limit on the Gunn-Peterson  optical depth
following Songaila et al.~(1999).
Figure 3 shows the high-resolution echellette spectrum of SDSS 1044--0125 in
the  Ly$\alpha$ forest region (binned to 2 \AA/pixel). 
The continuum level is approximated
by a $\nu^{-0.5}$ power law as above.
Even the most transparent part of the forest does not return
close to the continuum level at this resolution.
It is evident that the  Ly$\alpha$ forest is much stronger in SDSS 1044--0125
than in SDSSp J033829.31+002156.3 (Figure 2 of \cite{S99}),
where a fraction of the forest has flux comparable to the extrapolated
continuum.
In the region between 7926 \AA\ and
7929 \AA, SDSS 1044--0125 has an optical depth $\tau = 0.35 \pm 0.07$,
where the error bar only reflects the statistical noise in the spectrum
(note the noise level indicated in the figure).
This value changes to 0.40 and 0.31 for power law slopes of
0.0 and $-1.0$, respectively.
Therefore, we adopt a conservative limit of $\tau < 0.5$ at this redshift
of 5.52.
With higher resolution and signal-to-noise ratio, we might be able to
select regions even less affected by the Ly$\alpha$ forest lines.
Therefore, it is only an upper limit.
For comparison, Songaila et al.~derived  $\tau < 0.1$ for $z=4.72$
with similar resolution.

The Gunn-Peterson effect analysis above is based on an attempt to measure 
the amount of flux between Ly$\alpha$ forest lines (e.g., \cite{G94}).
At redshift higher than five, even with a moderately high resolution spectrum,
these forest lines overlap, making it impossible to find a
truly ``line-free'' region.
Modern hydrodynamic simulations and semi-analytic 
models show that under the influence of gravity, 
the intergalactic medium becomes clumpy, and the Gunn-Peterson optical depth
should vary even in the lowest column density regions (e.g. \cite{Bi92},
\cite{Jordi93}, \cite{Cen94}, \cite{Hernquist96}).
The minimum absorption regions in the forest merely represent regions
that are most underdense in this fluctuating Gunn-Peterson effect.
An accurate measurement of the Gunn-Peterson effect and the ionizing background 
from the high resolution spectrum of SDSS 1044--0125 requires detailed
comparison with cosmological simulations; this is beyond the scope
of the current paper.

Figure 2  also shows the detection of an intervening MgII absorption system.
The MgII doublet $\lambda 2796.4+2803.5$ is detected at wavelengths
9166.7 \AA\ and 9190.3 \AA\ in the high-resolution spectrum; 
the redshift of this system is $z_{abs}=2.278$.
The rest frame equivalent widths of the doublet lines are 2.43 and 1.90 \AA,
respectively. This system is very similar to the one detected in 
SDSSp J033829.31+002156.3 ($z_{abs}=2.304$, \cite{S99}).
It is possible that SDSS 1044--0125 is amplified by lensing from this intervening
absorber. However, we saw in \S~2 that this object is {\em unresolved}
under $0.4''$ seeing in the K band. 

\subsection{Number Density of Very High Redshift Quasars}

The total area of SDSS imaging data that we have searched for high-redshift
quasars thus far is of order 600 deg$^2$. 
All 
that satisfy
$z^*<19.3$ and $i^* - z^* > 2.2$ in this 600 deg$^2$ region have
been observed spectroscopically. Only SDSS 1044--0125 is identified as
a high-redshift quasar; 
the remaining objects are L and T dwarfs.
Using the luminosity function and redshift dependence of \cite{SSG95} (for the $\Omega=1$ model),
extrapolating it to higher redshifts and assuming $f_{\nu} \propto \nu^{-0.5}$,
we predict that in a total area of 600 deg$^2$, 
for $z>5.65$ ($i'$-dropout objects ), there should be 1.5 quasars with $M_{1450} < -27.0$
and 1.1 quasars with $M_{1450} < -27.2$.
For $z>5.8$, the extrapolation predicts 1.1 and 0.8 quasars for
$M_{1450} < -27.0$ and $-27.2$, respectively.
This assumes that our selection efficiency is 100\%.
The Schmidt-Schneider-Gunn luminosity function is derived using
objects with $2.7 < z < 4.7$ and
$-27.5 < M_{B} < -25.5$.
Although it is difficult to draw any reliable conclusion from the
observation of a single high-redshift quasar,
the discovery of SDSS 1044--0025 is consistent with the
expectations from this rather large extrapolation from lower redshift
results. 
Assuming that the same luminosity function holds at even higher redshift,
the SDSS will be able to discover
one quasar at $z\gtrsim 6$, $z^* \lesssim 19$ in every 1500 deg$^2$ of the survey.

SDSS 1044-0125 is a very luminous quasar.  Assuming that (1) its bolometric
luminosity equals the Eddington luminosity, $L_{\rm bol}=L_{\rm Edd}=1.5\times
10^{38}(M_{\rm BH}/{\rm M_\odot})~{\rm erg~s^{-1}}$; (2) its intrinsic
continuum spectrum is the same as the mean spectral template of \cite{Elvis94},
and (3) neither beaming nor lensing affects the
observed flux, we find a black hole mass of $M_{\rm BH}=3.4\times 10^{9}{\rm
M_\odot}$ in the $\Lambda$--model, or $M_{\rm BH}=2.7\times 10^{9}{\rm
M_\odot}$ in the $\Omega=1$ model.   In either case, the implied black hole mass
is quite large, similar to that of the black hole at the center of the
nearby giant elliptical galaxy M87 (Harms et al.~1994, Macchetto
et al.~1997). 
If the quasar were radiating below the Eddington limit,
the inferred black hole mass would be even higher.
In the \cite{Elvis94} template, $\approx
1\%$ of the bolometric luminosity is emitted in the observed $z^\prime$ band.  If this
fraction is larger for SDSS 1044-0125, 
the implied black hole mass would be reduced.
Note that the universe was less than 1 Gyr old at this redshift, while the
Eddington time scale, the $e$-folding time for the growing of a black hole
shining at the  Eddington luminosity, is $4 \times 10^7 (\epsilon/0.1)$ yr,
where $\epsilon$ is the radiative efficiency for the accretion.
If the black hole started accreting with an initial mass of 
$\sim 10^3 \rm M_\odot$ with 10\% efficiency,
the seed black hole would have to form and begin accreting
at redshift well beyond 10 in order to grow to $3\times 10^{9} \rm M_\odot$
at $z=5.80$ (see also \cite{Turner91}).
Forming such a massive black hole in such a short time is a remarkable
feat.
The observations of high-redshift quasars can be used to constrain the formation
epoch of the first star clusters and the fueling process of early
black holes.

How likely is it to find a quasar like SDSS
1044-0125 in popular cold dark matter cosmological models?  
We use the following simple model to estimate the abundance of high-redshift
quasars (see also \cite{HH00}):
\cite{Magoo98} have found a
correlation between central black hole mass and bulge mass, $M_{\rm BH}/M_{\rm
bulge}=6\times10^{-3}$ in nearby galaxies.  If this correlation holds at high
redshift, this would imply a bulge mass of $5.7\times10^{11}~{\rm M_\odot}$
($\Lambda$-model) or $4.5\times10^{11}~{\rm M_\odot}$ ($\Omega=1$) for the host
galaxy of SDSS 1044-0125, and a lower limit of $5.7\times10^{12}~{\rm M_\odot}$
($\Lambda$-model) or $4.5\times10^{12}~{\rm M_\odot}$ ($\Omega=1$) for its dark
halo, assuming $M_{\rm halo}/M_{\rm bulge}\geq \Omega_{\rm DM}/\Omega_{\rm
b}\approx 10$.  

The comoving abundance of dark matter halos at this epoch is
very sensitive to this halo mass, and can be estimated by
means of the \cite{PS74} formalism.  In a $\Lambda$-CDM  model, 
assuming $\sigma_{8} = 0.87$ and an untilted primordial power spectrum,
for parent
halo masses of $10^{12}$ M$_\odot$, $6 \times 10^{12}$ M$_\odot$,
and $10^{13}$ M$_\odot$, we expect  50,000, 60, and 4  candidate halos
respectively 
within the survey volume in a redshift window $\Delta z = 1$.  
The duty cycle of quasar activity is poorly known, but is
certainly much less than unity (\cite{HH00}).  Given these
uncertainties, this model is not contradictory to our discovery of a
single quasar like SDSS 1044-0125 in the 600 deg$^2$ survey area.

Note that this model does not address the physical process by which the massive
black hole formed.
The assumptions we made about $M_{\rm BH}/M_{\rm bulge}$  and the
lifetime of the quasars are completely untested at high redshift.
Indeed, Rix et al.~(1999) argue that the host galaxies of $z \sim 2$
quasars are appreciably less luminous than the universal $M_{\rm BH}/M_{\rm
bulge}$ hypothesis would imply.  
Because of its
high luminosity, SDSS 1044-0125 likely probes the exponential, high--mass tail
of the underlying dark halo distribution, making predictions of the expected
number counts also sensitive to cosmological parameters, especially the
normalization of the power spectrum (i.e. $\sigma_8$).  In principle, the SDSS
survey will be able to probe quasars $\approx 1$ mag fainter than SDSS
1044-0125, and could reveal tens of additional sources at $z\approx 6$.  
The detection of these objects will yield strong constraints on cosmological models
for the formation and evolution of quasars at very high redshifts.

\bigskip
The Sloan Digital Sky Survey (SDSS) is a joint project of the
University of Chicago, Fermilab, the Institute for Advanced Study, the
Japan Participation Group, The Johns Hopkins University, the
Max-Planck-Institute for Astronomy, Princeton University, the United
States Naval Observatory, and the University of Washington.  Apache
Point Observatory, site of the SDSS, is operated by the Astrophysical
Research Consortium.  Funding for the project has been provided by the
Alfred P. Sloan Foundation, the SDSS member institutions, the National
Aeronautics and Space Administration, the
National Science Foundation, the U.S. Department of Energy, and 
Monbusho, Japan.
The SDSS Web site is {\tt http://www.sdss.org/}.
This publication makes use of data products from the Two Micron All Sky Survey, which is a
joint project of the University of Massachusetts and the Infrared Processing and Analysis
Center/California Institute of Technology, funded by NASA and NSF.
XF and MAS acknowledge additional
support from Research Corporation, NSF grant AST96-16901, the
Princeton University Research Board, and a Porter O. Jacobus Fellowship.
RHB acknowledges support from the Institute of Geophysics and Planetary Physics 
(operated under the auspices of the U.S. Department of Energy by 
the University of California
Lawrence Livermore National Laboratory under contract No.~W-7405-Eng-48).
ZH acknowledges support from Hubble Fellowship grant
HF-01119.01-99A.
DPS acknowledges support from NSF grant AST99-00703.
We thank Wolfgang Voges, Hans-Walter Rix, David Weinberg, and  Peng Oh for helpful comments, and
the expert assistance of Bob Goodrich and Terry McDonald during the Keck observations.

\begin{deluxetable}{rrrrrr}
\tablenum{1}
\tablecolumns{7}
\tablecaption{SDSS and K-band Photometry of SDSSp J10:44:33.04 --01:25:02.2}
\tablehead
{
$u^*$ & $g^*$ & $r^*$ & $i^*$ & $z^*$ & K
}
\startdata
22.91 & 23.91 & 25.13 & 21.81 & 19.23 & 17.02\\
$\pm$ 0.53 & $\pm$ 0.49 & $\pm$ 0.62 & $\pm$ 0.19 & $\pm$ 0.07 &
$\pm$ 0.04\enddata

\tablenotetext{}{The SDSS photometry ($u^*,g^*,r^*,i^*,z^*$) is
reported in terms of {\em asinh magnitudes} on the AB system.
The asinh magnitude system is defined in Lupton, Gunn \& Szalay (1999);
it becomes a linear scale in flux when the absolute value of the
signal-to-noise ratio is less than about 5. In this
system, zero flux corresponds to 24.24, 24.91, 24.53, 23.89, and
22.47, in $u^*$, $g^*$, $r^*$, $i^*$, and $z^*$, respectively; larger
magnitudes refer to negative flux values. The K band photometry is on
the Vega-based system.  $E(B-V)$ in this direction is 0.054, from Schlegel et al. 1998}
\end{deluxetable}

\begin{deluxetable}{cccc}
\tablenum{2}
\tablecolumns{4}
\tablecaption{Emission Line Properties}
\tablehead
{
Line & $\lambda$ (\AA) & redshift & EW (\AA, rest frame) 	
}
\startdata
OI+SiII 1302 & 8858.5 $\pm$ 3.8 & $5.802 \pm 0.003$ & 3.4 $\pm$ 0.5 \\
SiIV+OIV] 1400 & 9507.3 $\pm$ 3.7 & $5.789 \pm 0.003$ & 7.0 $\pm$ 0.4
\enddata
\end{deluxetable}

%
\newpage
\begin{figure}
\vspace{0.5cm}

\epsfysize=600pt \epsfbox{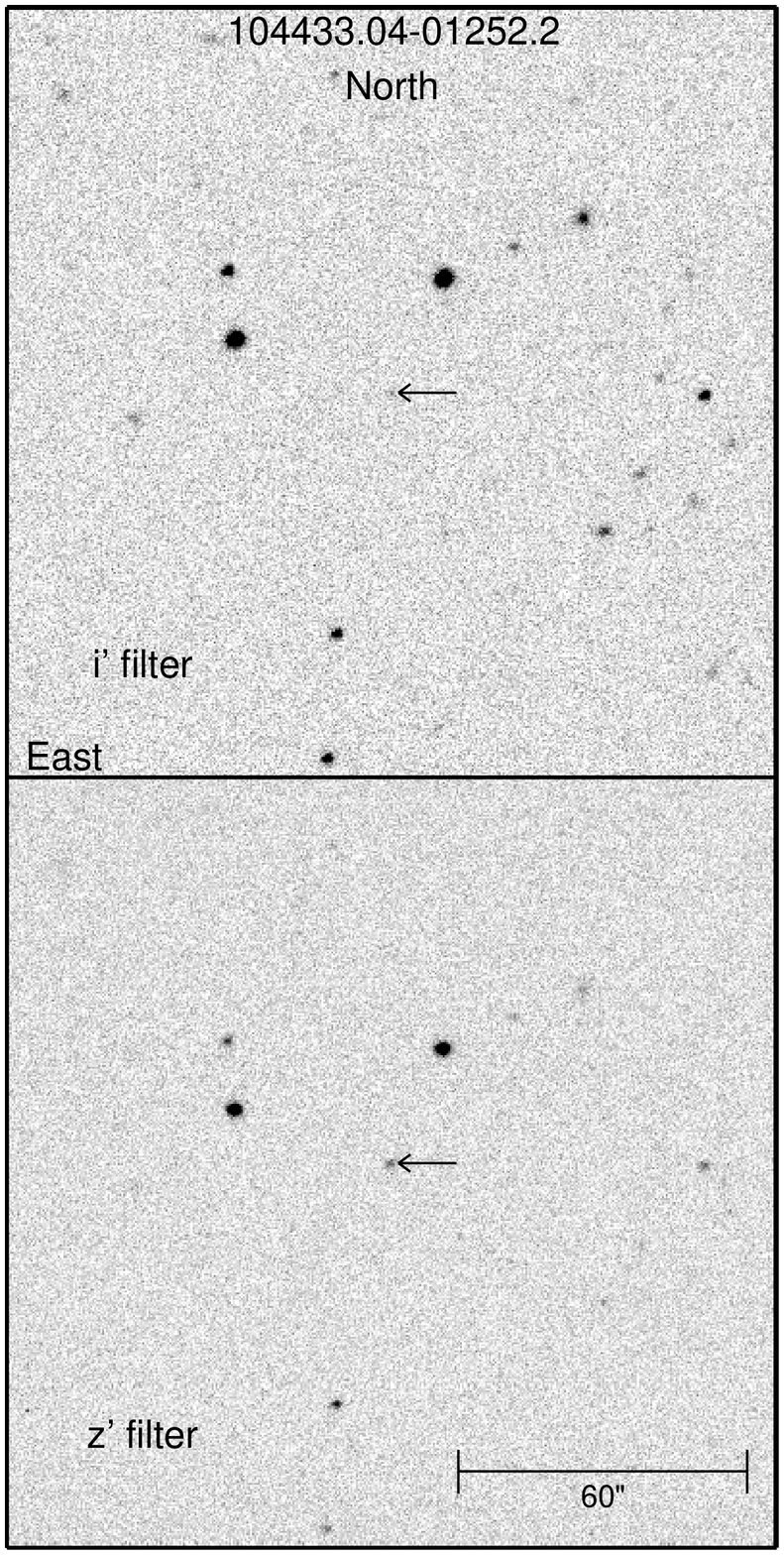}

\vspace{0.5cm}
Figure 1. Finding chart for SDSS 1044--0125 (discovery image from the SDSS).
The field is $160''$ on a side. The field is given in both the $i'$ and $z'$
bands (54.1s exposure time).

\end{figure}
\begin{figure}
\vspace{-7cm}

\epsfysize=750pt \epsfbox{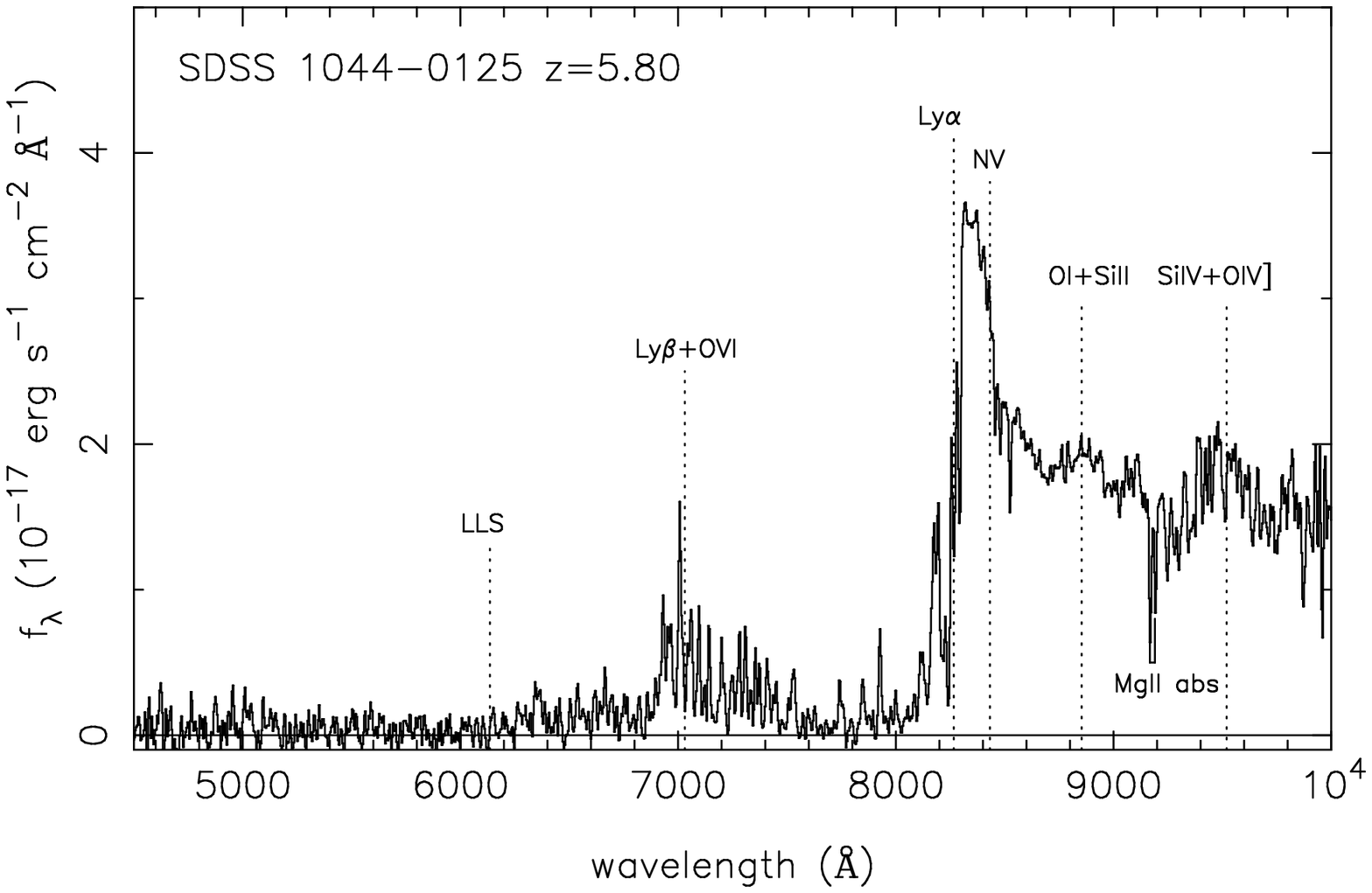}
Figure 2. Optical spectrum of SDSS 1044--0125 observed with KeckII/ESI.
The total exposure time is 3600s. The spectrum is smoothed to 
4 \AA/pixel. The spectral resolution is $\sim 8$ \AA\ at $\lambda = 9000$\AA.

\end{figure}
\begin{figure}
\vspace{-7cm}

\epsfysize=750pt \epsfbox{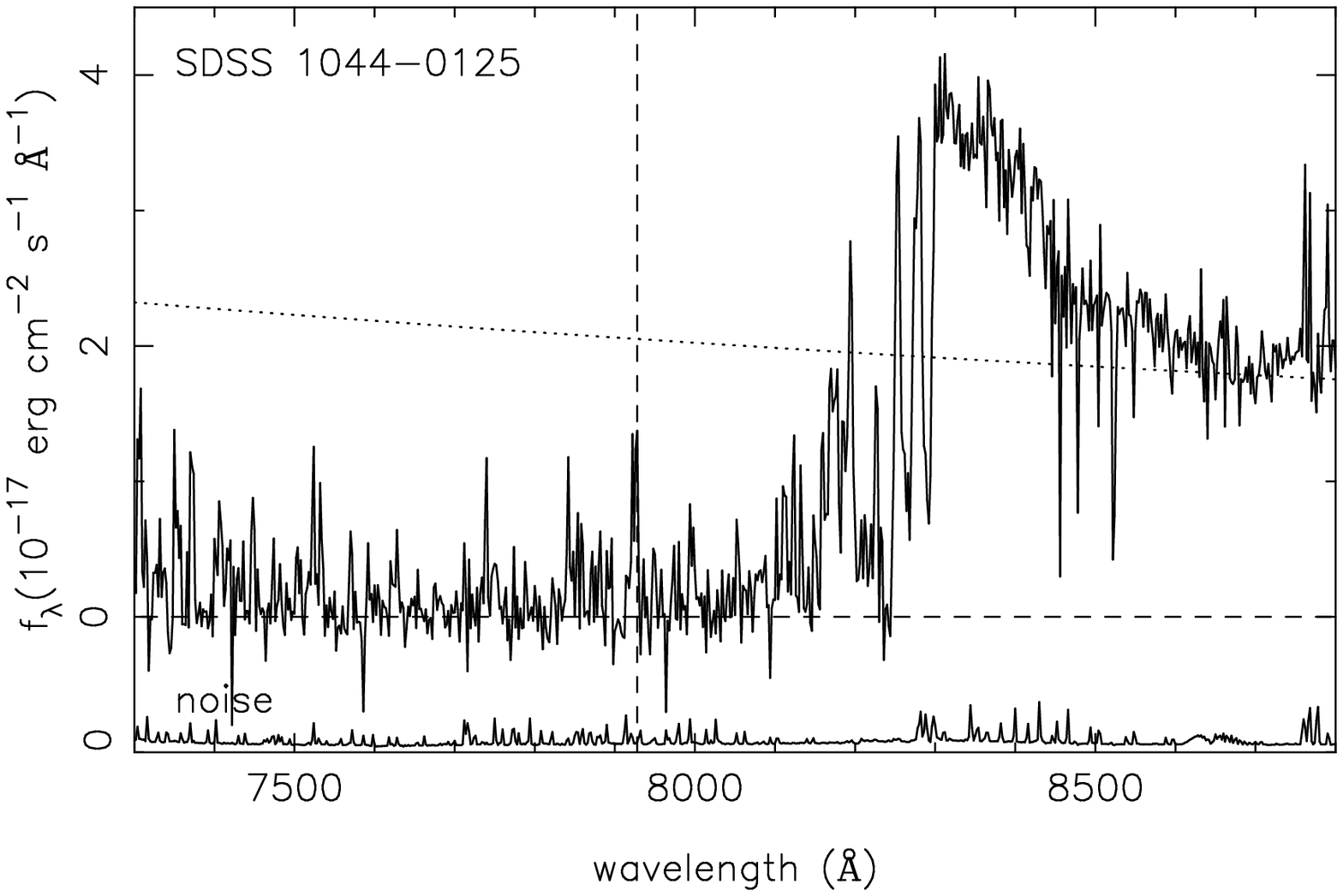}
Figure 3. ESI Echellete spectrum over the range 7300 -- 8800 \AA.
The spectrum has been smoothed to a resolution of 2 \AA.
The dotted line shows a $f_\nu \propto \nu^{-0.5}$ continuum, normalized to
the region 8650 -- 8730 \AA.
The dashed line indicates the region with minimum absorption
($\tau \sim 0.4$, see \S 3.1).
The lower panel shows the noise level.
\end{figure}

\end{document}